# Exchange-striction driven ultrafast nonthermal lattice dynamics in NiO

Y. W. Windsor[1], D. Zahn[1], R. Kamrla[2], J. Feldl[3], H. Seiler[1], C.-T. Chiang[2], M. Ramsteiner[3], W. Widdra[2], R. Ernstorfer[1], L. Rettig[1]

[1] Department of Physical Chemistry, Fritz Haber Institute of the Max Planck Society, Faradayweg 4-6, 14195 Berlin, Germany
[2] Institute of Physics, Martin-Luther-Universität Halle-Wittenberg, 06120 Halle, Germany
[3] Paul-Drude-Institut für Festkörperelektronik, Leibniz-Institut im Forschungsverbund Berlin e.V., Hausvogteiplatz 5-7, 10117 Berlin, Germany

**Abstract:**

We use femtosecond electron diffraction to study ultrafast lattice dynamics in the highly correlated antiferromagnetic (AF) semiconductor NiO. Using the scattering vector (Q) dependence of Bragg diffraction, we introduce a Q-resolved effective lattice temperature, and identify a nonthermal lattice state with preferential displacement of O compared to Ni ions, which occurs within ~0.3 ps and persists for 25 ps. We associate this with transient changes to the AF exchange striction-induced lattice distortion, supported by the observation of a transient Q-asymmetry of Friedel pairs. Our observation highlights the role of spin-lattice coupling in routes towards ultrafast control of spin order.

NiO has been of interest for several decades, both from fundamental and application perspectives [1–4]. Due to strong correlations, ab-initio descriptions of this large band gap charge-transfer semiconductor are challenging [5–12]. The open Ni *d*-shell and strong superexchange leads to high-temperature antiferromagnetic (AF) order ($T_N$ ≈ 523 K), making NiO a promising candidate for room-temperature spintronic applications [13–15]. In this context, pioneering experiments have demonstrated coherent excitation of high-frequency magnon modes by THz pulses [16–18], fueling the promise of ultrafast AF spintronics. For such purposes, understanding energy transport due to coupling between the material's subsystems is of importance, in particular those to magnetic order. In equilibrium, NiO exhibits strong spin-lattice coupling and exchange striction [19], leading to a rhombohedral lattice distortion (RLD) along the [111] direction of its nominally cubic structure below $T_N$ [20]. Spin-phonon coupling in NiO has been recently studied in the context of magnon damping in devices [21]. However, energy transfer dynamics and couplings between the various subsystems upon optical excitation have been little studied so far [22]. In particular, to date there is no account of ultrafast lattice dynamics in NiO.

In the presence of a band gap, optically excited carriers can radiatively decay, and they can transfer energy to another subsystem, e.g. the lattice. This is typically described using coupled-heat-baths models [23–25], where the subsystems' transient temperatures are described by coupled rate equations. While often successfully employed [26,27], an implicit assumption is that the baths themselves remain thermalized, such that the carriers and the phonons always follow Fermi-Dirac and Bose-Einstein distributions. Whereas this is considered valid for many metals because of homogenous electron-phonon coupling and rapid electronic thermalization, this assumption has been recently challenged even for simple metallic systems [28–31]. In semiconductors, electron-phonon coupling is often strongly heterogenous [32], and the phonon dispersion is often more complicated than in metals. Consequently, photoexcited semiconductors may experience a prolonged nonthermal lattice state, in which unconventional relaxation processes may occur. To date only a few reports describing nonthermal lattice dynamics of semiconductors have been published [33–37], and details about the underlying microscopic processes are scarce. In complex materials such as NiO, lattice dynamics may be substantially influenced by effects beyond electron-phonon coupling, such as spin-lattice coupling via exchange striction.

Here we use femtosecond electron diffraction (FED) to study photoinduced lattice dynamics in NiO. Variants of FED have recently been used to study lattice dynamics in several other systems, including metals [38,39], semiconductors [34,40], heterostructures [41,42], and systems involving structural phase transitions [43,44]. We present an approach to characterize the transient lattice state, by converting Bragg reflection intensities into units of Kelvin, producing a series of temperatures associated with each scattering vector of the probe electrons (*Q*). Through analysis of these Q-dependent effective lattice temperatures we identify a strongly non-thermal lattice state after excitation, which lasts for tens of picoseconds. Employing the Q-dependent $Ni^{2+}$ and $O^{2-}$ scattering factors we gain a degree of element-sensitivity, and find that the lattice response to photoexcitation is primarily on oxygen ions, and less on Nickel. Based on the initial energy transfer time scale of ~0.3 ps, we propose a scenario in which photoexcitation perturbs the antiferromagnetically-induced RLD. This is supported by photoinduced changes in observed Q values that occur due to changes in the shape of the unit cell (i.e. the RLD), suggesting that the observed nonthermal lattice state originates from preferential occupation of phonon modes associated with the RLD.

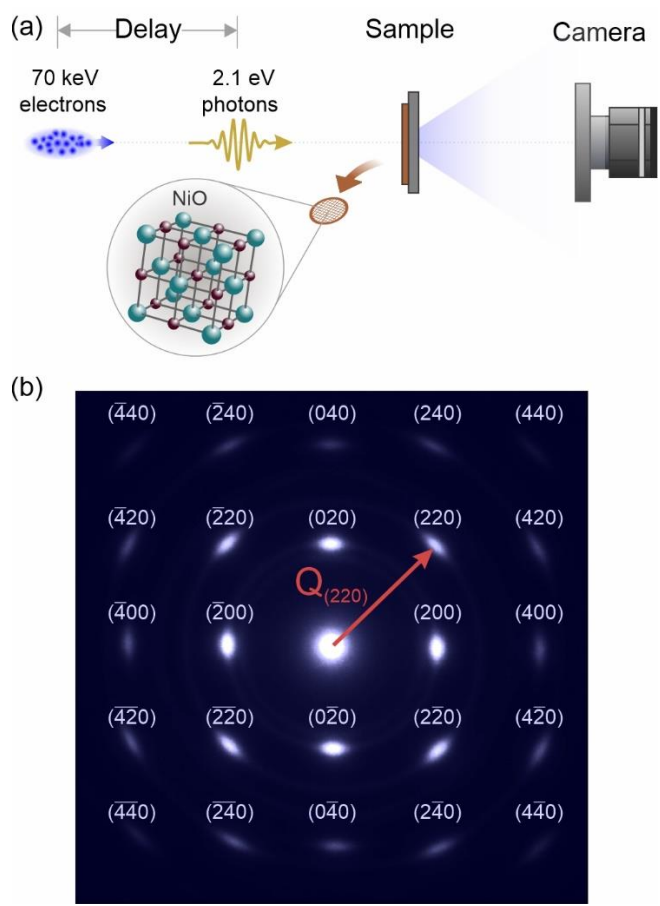

**Fig. 1** – (a) Experimental scheme. (b) Example diffraction pattern. Arrow: exemplary scattering vector. In this work "Q" denotes the length of this vector.

We use a 20 nm thick single crystal of NiO, epitaxially grown on a [001]-oriented NaCl crystal, which was subsequently dissolved. Diffraction and Raman measurements confirmed its bulk-like properties (see supplement). Experiments were conducted at room temperature ($T_0$) using a compact setup [45] (Fig. 1a), with femtosecond laser excitation of $h\nu = 2.16$ eV and 5.2±1.3 mJ cm$^{-2}$ incident fluence. This photon energy slightly exceeds half the charge-transfer gap of $\Delta \approx 3.8$ eV [22,46,47], and two-photon absorption is likely dominant over linear absorption (see supplement). 70 keV probe electrons transmit through the sample, producing patterns as in Fig. 1b. The response function is estimated at 200 fs.

The lattice response was measured using diffraction patterns from different pump-probe delays $t$. The observed diffraction spots correspond to reflections with Miller indices ($hk0$). Faint rings are also observed, all corresponding to NiO Bragg reflections, likely from polycrystalline regions. Intensities of >50 diffraction spots were extracted, covering 10 values of $Q$ (defined as $Q = 2\lambda^{-1} \sin\theta$ ; $\lambda$ and $\theta$ are electron wavelength and Bragg angle; see Fig. 1b). Intensities from spots of equal $Q$ were averaged, producing 10 independent intensity observables $I_Q(t)$, shown in Fig. 2a. All $I_Q(t)$ curves exhibit an initial sub-picosecond drop, followed by a slower process.

We describe Bragg intensities using structure factor calculations. Intensities of ($hk0$) reflections depend only on the size of $Q$[i]:

$$I_Q \propto \left| f_{Ni}(Q) e^{-Q^2 B_{Ni}} + f_O(Q) e^{-Q^2 B_O} \right|^2 \quad (1)$$

Here $f_{Ni}$ and $f_O$ are the scattering factors of the Ni$^{2+}$ and O$^{2-}$ ions, which are tabulated functions of $Q$ [48]. $B_{Ni}$ and $B_O$ are the Debye-Waller (DW) factors for Ni and O ions, which are tabulated as functions of temperature for NiO [48]. DW factors can be expressed as $B_x = \frac{2}{3}\pi^2 \langle u_x^2 \rangle$ ($x$ represents Ni or O), in which $\langle u_x^2 \rangle$ are the *time-averaged mean square displacement* (MSD) of the Ni$^{2+}$ or O$^{2-}$ ions, i.e. a measure of each atom species' vibrations $u_x(t)$ about their mean position. Upon heating in equilibrium conditions, the MSDs and thereby the DW factors increase due to the growing phonon population, causing a reduction in diffracted intensity (DW effect). Similarly, MSDs can also increase upon photoexcitation if a phonon population is induced, and serve as reliable measures for the lattice response to laser excitation [40].

---

[i] The rhombohedral distortion away from 90° reaches ~0.06° at room temperature, so a cubic structure is commonly assumed.

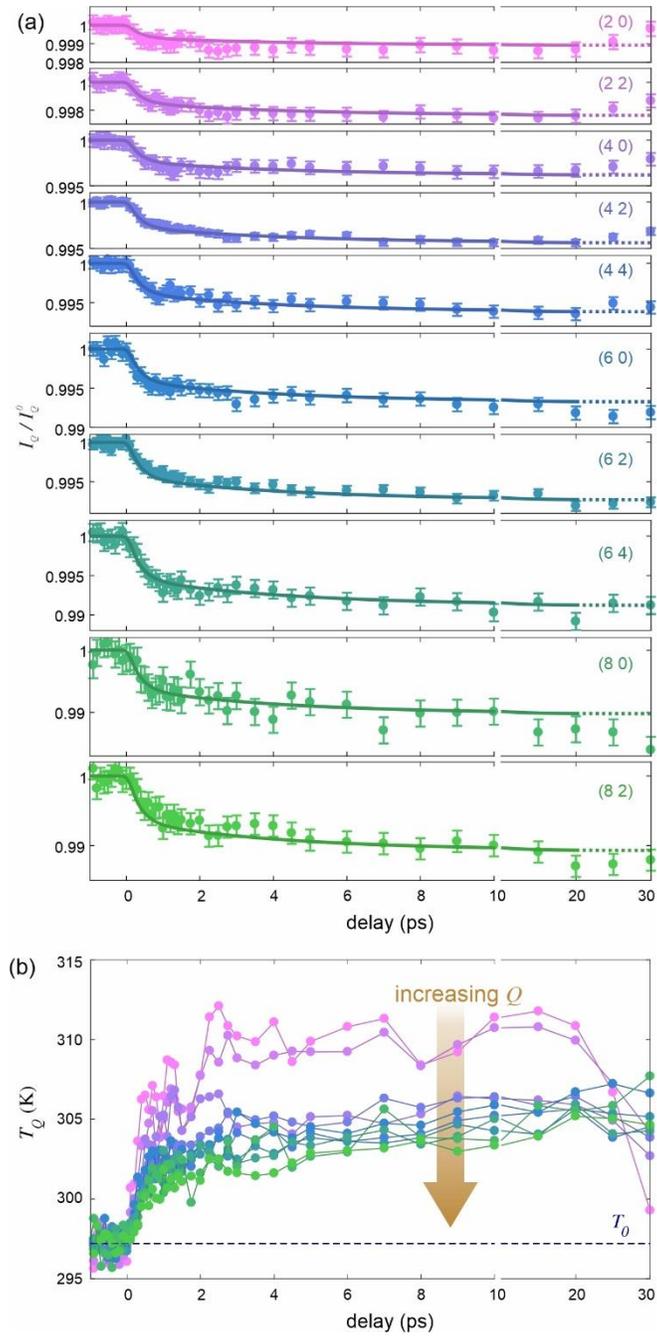

**Fig. 2** – (a) Normalized transient Bragg intensities: each $I_Q$ curve represents the mean response of all Bragg reflections with the same scattering vector length $Q$. Labels: representative $(hk0)$ indices. Lines are calculated from the $\bar{T}$ fit (Fig. 4b) using Eq. 1. $I_Q^0$ is the unpumped $I_Q$. (b) $Q$-dependent effective lattice temperatures $T_Q$. Each curve is calculated from Fig. 2a using tabulated temperature-dependent DW factors in Eq. (1).

Since $B_{Ni}$ and $B_O$ in Eq. (1) cannot be analytically separated, we adopt a temperature-based approach. We insert the tabulated $B_{Ni}(T)$ and $B_O(T)$ into Eq. (1) to convert the relative intensities $I_Q(t)$ into temperatures $T_Q(t)$, in units of Kelvin. The $T_Q$ curves, shown in Fig. 2b, represent *effective* transient lattice temperatures, because our conversion is based on equilibrium (i.e. *thermal*) DW factors, while the phonon population may be *nonthermal* after photoexcitation. Nevertheless, the $T_Q(t)$ curves provide a Q-resolved picture which can now be used to describe the nonthermal state of the lattice.

The $T_Q$ curves in Fig. 2b deviate significantly from each other, both in the rise magnitude (the lower $Q$ curves reach higher temperatures) and in their qualitative behavior. To explain this, we consider that different $Q$ provide varying sensitivity to the $Ni^{2+}$ and $O^{2-}$ ions through the Q-dependence of their respective scattering factors $f_{Ni}$ and $f_O$ in Eq. (1). Fig. 3a presents their ratio $\eta(Q) = f_{Ni}/f_O$, in a range covering all probed $Q$ values. Due to the large difference between scattering from $Ni^{2+}$ and from $O^{2-}$, $\eta$ shows a significant variation in this range, demonstrating that Bragg reflections at higher $Q$ values are more sensitive to scattering from $Ni^{2+}$ than from $O^{2-}$.

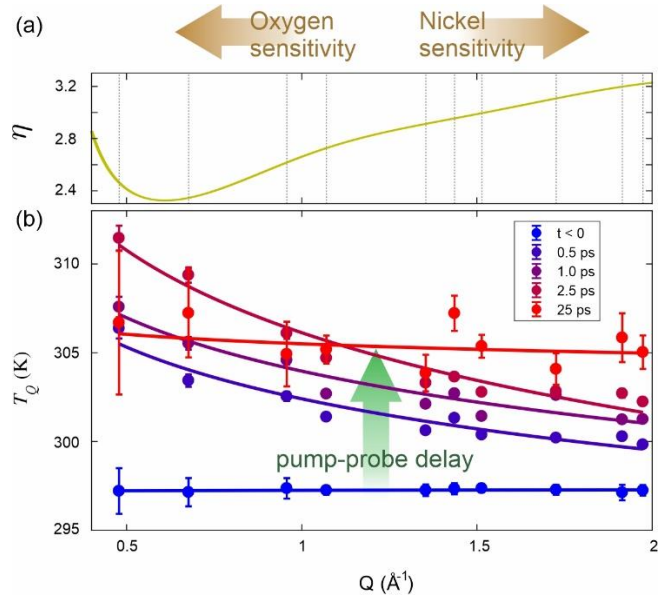

**Fig. 3** – (a) Calculated ratio between $Ni^{2+}$ and $O^{2-}$ scattering factors $\eta(Q) = f_{Ni}/f_O$ as function scattering vector length $Q$. Dashed lines mark discreet $Q$ values probed in the experiment. (b) Q-dependent lattice temperatures $T_Q$ at selected delays. Solid lines are fits to $T_Q \propto Q^b$ .

To demonstrate this, Fig. 3b presents $T_Q$ as a function of $Q$ at selected delays. Before excitation $T_Q$ exhibits no $Q$ dependence, as expected in a thermal state (all $T_Q = T_0$). After excitation, $T_Q$ exhibits an overall increase in temperature due to the excitation, with a *continuous* reduction upon increasing $Q$. This marks a departure from equilibrium behavior. Given the higher sensitivity to oxygen vibrations at lower $Q$ (Fig. 3a), this result indicates that the MSD of oxygen $\langle u_O^2 \rangle$ initially grows *disproportionately* more than that of nickel $\langle u_{Ni}^2 \rangle$, demonstrating a nonthermal lattice state at early times. This trend is subsequently suppressed, such that at 25 ps the $Q$-dependence of $T_Q$ is nearly flat, indicating thermalization of the lattice. This disproportionate growth is not to be confused with the *absolute* difference between O and Ni MSDs, which differ also in equilibrium [49].

We quantify this nonthermal disproportionality between the O and Ni vibrational responses by empirically describing these $Q$-dependences as $T_Q \propto Q^b$ (solid lines, Fig. 3b). Fig. 4a presents $b(t)$ at each delay, exhibiting an initial sub-picosecond reduction, and reaching a minimum at 3 ps, corresponding to the most pronounced disproportionality between O and Ni. After ~15 ps, $b(t)$ recovers toward 0, i.e. to a thermal lattice state, which is also reflected in the difference between the $T_Q$ curves in Fig. 2b.

$\eta(Q)$ also enables a reliable description of the *mean* lattice response. Averaging $\eta$-weighted $T_Q$ data (Fig. 2b) produces a mean lattice response $\bar{T}(t)$, presented in Fig. 4b. As in $I_Q(t)$, $\bar{T}(t)$ exhibits a two-step response, well-described by a biexponential. The best fit (solid line) yields a sub-picosecond process ($\tau_1 = 0.31 \pm 0.08$ ps) and a second, slower one ($\tau_2 = 4.1 \pm 1.3$ ps), see dashed lines. The total temperature rise is $\Delta T = 9.6 \pm 0.7\ K$, of which 60% is the fast process. Using the NiO lattice heat capacity [50], we find that $\Delta T$ corresponds to a maximal change of $16 \pm 1$ meV/unit cell in the lattice energy density. Careful evaluation of the data concluded that the fast process is intrinsic, while the slower process evolves as the measurement progresses (see supplement). Late delays were omitted from the fit because they exhibit recovery, similar to that of $b(t)$ (Fig. 4a). We convert $\bar{T}(t)$ back into Bragg intensities by $\eta$-weighing the $\bar{T}(t)$ fit and plugging it into Eq. 1. This produces the lines in Fig. 2a, in reasonably good agreement with the data. Disagreements exist because this washes out the distinctly nonthermal description in Fig. 4a. Our fit does not consider recovery, so at late delays the $Q$-dependence of $\eta$ causes $I_Q$ data with low/high $Q$ to be above/below the lines.

To interpret the stronger response of the O ions, we inspect the phonon dispersion of NiO. This reveals several optical phonons that preferentially move the O ions at the Brillouin zone boundaries (e.g. along [111]), in particular the LO' mode [21]. Below $T_N$, the optical modes' frequencies deviate significantly from the expected temperature dependence of crystal lattice anharmonicity [21], and they split in energy [51] due to the RLD [52,53]. Preferential coupling to such modes would lead to their enhanced population in a nonthermal phonon distribution, followed by thermalization of the excited phonon population to lower energy/momentum modes, on phonon-phonon scattering timescales [32,54]. A scenario leading to a preferential excitation of such modes could be a perturbation of the AF-induced RLD, for which a full crystallographic account was only recently reported [55]. Following Uchiyama [53], two effects contribute to this distortion, which acts along [111] (sketched in Fig. 4c). The first is a distortion caused directly by the nearest-neighbor superexchange [56]. The second originates from Coulomb forces induced by an asymmetric charge distribution around the ions, due to AF-induced band folding [52,53], predicted to be the dominant contribution [53].

This suggests two possibilities for perturbing the distortion. The first is that the excitation weakens the AF order, and therefore also the charge asymmetry, both of which then weaken the distortion. The lattice response time ($\tau_1 \approx 0.31$ ps) should then reflect magnetic excitation times. Reported optically-excited magnon data indeed suggest similar times scales [57–59]. However, a magnetic diffraction experiment has disputed these results [60]. The second possibility is that the charge asymmetry is directly perturbed, without involving magnetism. An excitation above the gap could cause a displacive phonon excitation. However, time scales associated with electronic excitations in NiO are much shorter than $\tau_1$ [61], and displacive phonon excitations in similar binary oxides occurs on much shorter time scales [62,63], so we conclude that this possibility is unlikely.

This brings about the following scenario. Electrons are optically excited above the gap and magnetic order rapidly weakens. This reduces the asymmetric charge distribution, triggering lattice motions that weaken the RLD. Due to this preferential electron-phonon coupling, modes that lead to a reduction of the RLD are preferentially occupied, which is observed as disproportionally higher growth of $\langle u_O^2 \rangle$. The lattice subsequently reaches an elevated thermal state via phonon-phonon coupling within ~25 ps. This scenario should directly affect other observables, such as the phononic energy gap or the rhombohedral angle itself.

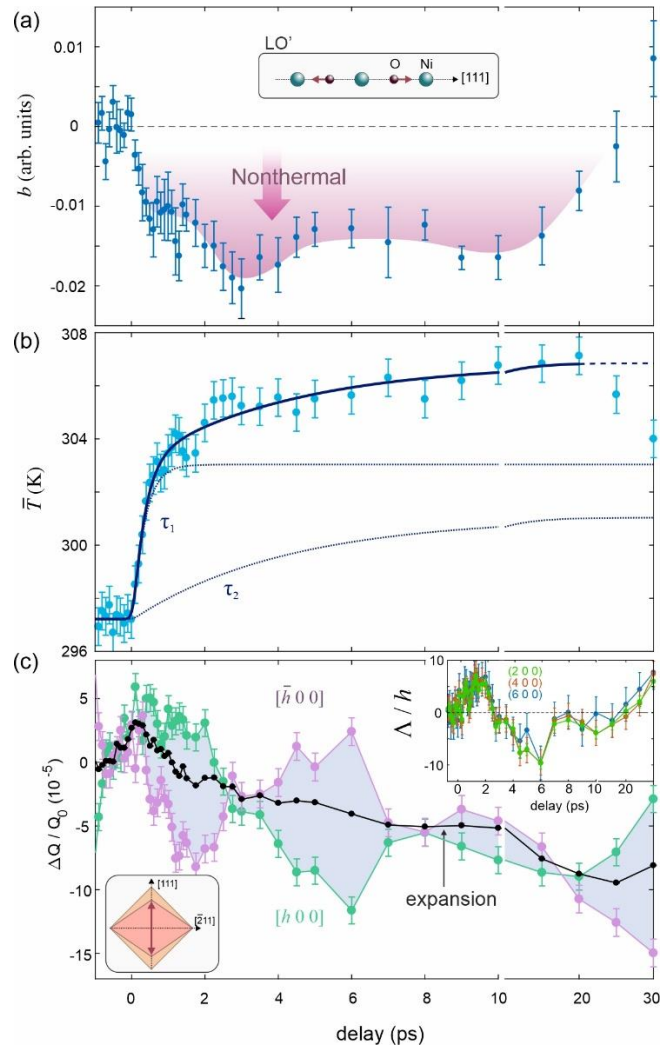

**Fig. 4** – (a) The coefficient $b(t)$, describing transient changes in the $Q$-dependence of $T_Q$ (as in Fig. 3b). The dashed line indicates zero (thermal lattice state). Inset: Schematic of the LO' motions (arrows) [21]. (b) $\eta$-weighted mean lattice temperature. The solid line is the best fit to a biexponential (convolved with Gaussian instrument response). Dashed lines represent the two individual processes. (c) Relative change in peak positions. Black: average of all $(h\,k\,0)$ reflections. The feature near time-zero is likely caused by space-charge interactions [64]. Green/Purple: evolution of $Q$ along two opposite directions: $[h\,0\,0]$ and $[\bar{h}\,0\,0]$ (averaged over $h$ = 2, 4, 6) Insets: Top: Asymmetry $\Lambda$ for $[h00]$ (in units of (detector pixels)$^2$), normalized by $h$. Bottom: sketch of the rhombohedral distortion of the cubic unit cell.

To support this scenario, we consider how varying the RLD would affect the observed scattering vectors, i.e. the *positions* of the spots (Fig. 1b). We divide them into Friedel pairs of the form $(hk0)$ and $(\bar{h}\bar{k}0)$, and extract the change in peak position $\Delta Q(t) = Q(t) - Q_0$ for each spot individually, as well as the average $\Delta Q$ of each pair. The average $\Delta Q$ curves of all pairs closely reproduce each other (black symbols in Fig. 4c), and exhibit a slow decrease of $Q$ indicating isotropic lattice expansion (i.e. *a = b* at all delays). Combining $\Delta Q$ with $\Delta T$ (Fig. 4b) produces an expansion coefficient of ~$10^{-5}$ K$^{-1}$, in agreement with literature [65]. However, the individual spots in every pair deviate symmetrically around this mean (shown for the $(h00)$ family in Fig. 4c). To quantify this, we introduce the asymmetry $\Lambda = Q^2(hk0) - Q^2(\bar{h}\bar{k}0)$ (accounting for Ewald's sphere curvature, see supplement). A nonzero $\Lambda$ represents *deviations from an orthonormal unit cell*, and scales as $\Lambda \propto (h+k)$ (see supplement). The inset presents $\Lambda(t)$ calculated from the same data as Fig. 4c, exhibiting this scaling. While $\Lambda(t)$ does not immediately translate into the magnitude of the RLD, its non-trivial dynamics demonstrate transient changes in it. Therefore, these data serve as direct evidence of a transient change in the *shape* of the unit cell upon excitation, supporting the scenario of a pump-induced changes in the RLD (lower inset, Fig. 4c). They underline the role of spin-lattice coupling through exchange striction in the ultrafast lattice dynamics of NiO. Ultimately, this efficient spin-lattice coupling may facilitate ultrafast spintronic applications e.g. by enabling structural control of magnetism.

In summary, we studied the lattice response of NiO to photoexcitation using femtosecond electron diffraction. Describing the data as scattering-vector-dependent effective lattice temperatures enabled us to study the transient nonthermal state of the lattice. Compared to thermal conditions, this state involves a disproportionally higher displacement of O ions compared to Ni. While the lattice response time is 0.31 ps, the nonthermal state persists for up to ~25 ps, after which the system reaches a new thermal state. We present a scenario in which this nonthermal state is facilitated by perturbation of the antiferromagnetically-induced rhombohedral lattice distortion. This is supported by observed changes in the asymmetry of Bragg peak positions of Friedel pairs, a hallmark of non-orthonormal systems. Our results shed light on the nature of the nonthermal lattice state in NiO, and demonstrate how spin-lattice coupling through exchange striction may play a key role in future ultrafast applications.

**Acknowledgements**

This work received funding from the DFG within the Emmy Noether program under Grant No. RE 3977/1, and within the Transregio TRR 227 Ultrafast Spin Dynamics (Projects A09, B07 and A06). Funding was also received from the Max Planck Society, and from the European Research Council (ERC) under the European Union's Horizon 2020 research and innovation program (Grant Agreement Number ERC-2015-CoG-682843). H.S. acknowledges support by the Swiss National Science Foundation under Grant No. P2SKP2.184100.

Ultrafast non-thermal lattice dynamics and spin-lattice relaxation in photoexcited nickel oxide
# Supplementary material

1. **Sample growth details**

Before deposition, the NiO growth rate was deduced from RHEED measured by deposition on a [001]-oriented Ag crystal. In the beginning of the deposition process, a [001]-oriented NaCl crystal was heated to 385 K to degas in UHV for 45 minutes. The cleanliness of the surfaces was confirmed by Auger electron spectroscopy at 335 K, chosen to prevent the crystal from charging. A 20nm NiO layer was then grown at room temperature on the [001] surface by evaporation of Ni in $10^{-6}$ mbar oxygen atmosphere. The crystal was then removed from vacuum, and the NaCl was dissolved in distilled water such that the NiO film remained flat on the water surface. The free-standing NiO crystal was then picked up using a Cu TEM grid with a spacing of 400 lines/inch.

All intensities and peak positions in our diffraction experiments were extracted from best fits to two-dimensional peak functions.

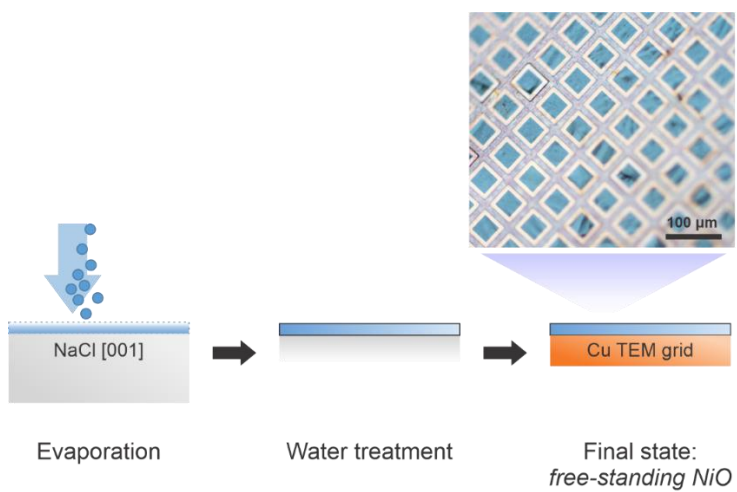

Figure S1 – schematic of the sample preparation process: NiO is evaporated on NaCl; the NaCl is dissolved using water, and the NiO film is placed on a Cu TEM Grid. Top: microscope image of the sample.

## 2. Raman Scattering

Raman spectroscopy measurements were performed to assess whether the sample's phonon dispersion agrees with that of bulk NiO. Measurements were conducted in the backscattering configuration with an optical excitation wavelength of 405 nm. The incident laser light was focused by a microscope objective onto the sample surface (spot diameter ~2 µm). The backscattered light was collected by the same objective (confocal configuration), spectrally dispersed by an 80 cm spectrograph (LabRam HR Evolution, Horiba/Jobin Yvon) and detected with a liquid nitrogen cooled CCD. For Rayleigh light suppression, a band pass filter with ultra-narrow spectral bandwidth was used. Figure S2 presents spectra taken at room temperature, with vertical lines indicating positions of Raman peaks due to second-order phonon scattering observed for bulk NiO.

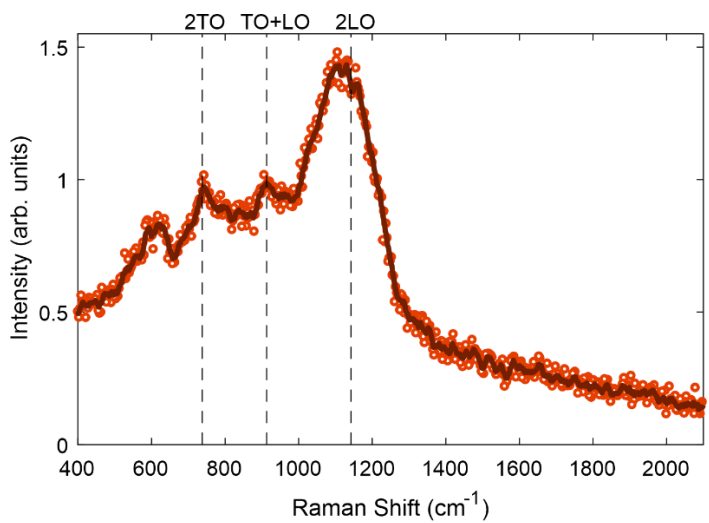

Figure S2 – Raman spectra collected from the sample at room temperature. Two-phonon features known from bulk NiO are indicated.

## 3. Photoexcitation

Here we detail the laser excitation. The probe beam was imaged on the sample, and its size was estimated using a 2000 lines/inch grid. We assume its profile $f_{probe}$ to be a flat top ellipsoid of full widths $(140 \pm 13 \times 114 \pm 13)\ \mu m^2$, and we label its perimeter as $P$ for convenience.

The pump beam was imaged at a point equivalent to the sample position. Its best fit to a two-dimensional Pseudo-Voigt profile $f_{pump}$ has a full width at half maximum (FWHM) of $(400 \pm 20 \times 245 \pm 12)\ \mu m^2$. The pump energy was $E_{pump} = 7.6 \pm 0.2\ \mu J$ per pulse, at a repetition rate of 4 kHz.

To estimate the incident fluence, we consider the fraction of this energy that overlaps with the probe pulse in an approach similar to the one presented by Harb et al. [1]. This is done by taking the ratio of two integrals over the pump profile: once only up to the limits of the probe pulse ($P$), and once over all space. The incident fleunce is then given by:

$$F_i = \frac{E_{pump}}{A_{probe}} \frac{\int_0^{2\pi} d\theta \int_0^{P(\theta)} r dr\, f_{pump}}{\int_0^{2\pi} d\theta \int_0^{\infty} r dr\, f_{pump}} (1-R) \tag{S1}$$

Here $A_{probe}$ is the probe spot area. $R$ is the reflectivity of the sample at the pump photon energy (2.16 eV), taken as $R=0.16$ [2]. Using Eq. (S1) we reach the final value of $F_i = 5.2 \pm 1.3$ mJ cm$^{-2}$.

Next we estimate the linear and two-photon absorption. The linear absorption coefficient at the pump photon energy is 260 cm$^{-1}$ [2]. Considering $F_i$, this produces a negligibly small absorbed fluence of $F_A^{1ph} = 0.003$ mJ cm$^{-2}$ (with ~20% error). For estimating two-photon absorption, we follow the steps described by Zeuschner et al. [3]. We make the same thin-layer approximation to reach an intensity of the form:

$$\frac{\partial I}{\partial z} = -\beta I^2 \quad \rightarrow \quad I(z) = \frac{I_0}{1 + \beta I_0 z} \approx I_0 - \beta I_0^2 z + O(z^2) \tag{S2}$$

Here z is the film thickness, $I_0$ is the incident intensity, and $\beta$ is the two-photon absorption coefficient. We also assume that the pump pulse is a Gaussian envelope with a FWHM of $\tau = 50 \pm 10$ fs. It then follows that the absorbed fluence from two-photon absorption is

$$F_A^{2ph} = \sqrt{\frac{\ln 4}{\pi}} \frac{F_i^2}{\tau} \beta z \tag{S3}$$

The available literature value for two-photon absorption at twice our pump energy is β = 0.12 cm/MW [4]. This produces $F_A^{2ph} = 0.091 \pm 0.047$ mJ cm$^{-2}$. This is over 30 times larger than the linearly absorbed fluence, indicating that the excitation is dominated by two photon absorption. These values take into account repeated internal reflections of the pump within the sample, which add ~4% more fluence. Importantly, we note that the value of $\beta$ was measured with 10 ns pulse lengths. As our pump pulse is ~50 fs, we regard $F_A^{2ph}$ as a lower limit for nonlinear absorption of our pump pulse.

## 4. Use of tabulated values

In this work we used tabulated values for the scattering factors and for the Debye-Waller factors. High-energy electron diffraction from NiO was studied in detail in Ref. [5]. The authors parametrized the scattering factors for $Ni^{2+}$ and $O^{2-}$ using a modified version of the Mott-Bethe formula of the form:

$$f(s) = \sum_{i=1}^{5} a_i \exp(-b_i s^2) + \frac{me^2}{2h^2} \frac{\Delta Z}{s^2} \tag{S4}$$

Here the X-ray scattering form factors are replaced by a parametrized sum of Gaussians, and the momentum transfer s is related to our definition of Q by $s = Q/2$. To Calculate $f_{Ni}(Q)$ and $f_O(Q)$, we used $\Delta Z = \pm 2$, which is the most conservative choice when considering the element sensitivity provided by $\eta$ (i.e. this value produces the smallest variation in $\eta$ across the measured Q range). We note that even when the Mott-Bethe formula [6,7] is used with tabulated X-ray scattering factors, the results of our analysis remain nearly unchanged. A more complete account of this approach was given by Peng in Ref. [8]. The Debye-Waller factors for the Ni and O ions in NiO were calculated as: (x is Ni or O) [9,10]

$$B_x = \sum_{n=0}^{4} a_n T^n \tag{S5}$$

## 5. Sample evolution during the measurement

In this section we discuss the evolution of the sample during the measurement, which could indicate degradation. The mean lattice temperature, $\bar{T}$, is presented in Fig. 4b of the main text. Its response to photoexcitation exhibits two processes: a sub-picosecond process and a second, longer process of a few picoseconds. The experimental data presented in this manuscript were accumulated over a period of ~24h. Figure S3a presents $\bar{T}$, as extracted at different stages of accumulation. Clearly the data spread improves with time. However, it is also apparent that the sub-picosecond process remains the same as time progresses (albeit better-resolved), while the slower process does not. The slower process is nearly absent at the beginning of the accumulation time, and grows in prominence as accumulation progresses. From this we conclude that the sub-picosecond process is intrinsic to the system, while the second process is induced during the experiment, such as due to increased absorption from photodoping.

The Q-dependent behavior described in the main text is not affected by this. To demonstrate this, Figure S3b presents the parameter $b(t)$ (as in Fig. 4a of the main text) at the same acquisition times as in a.

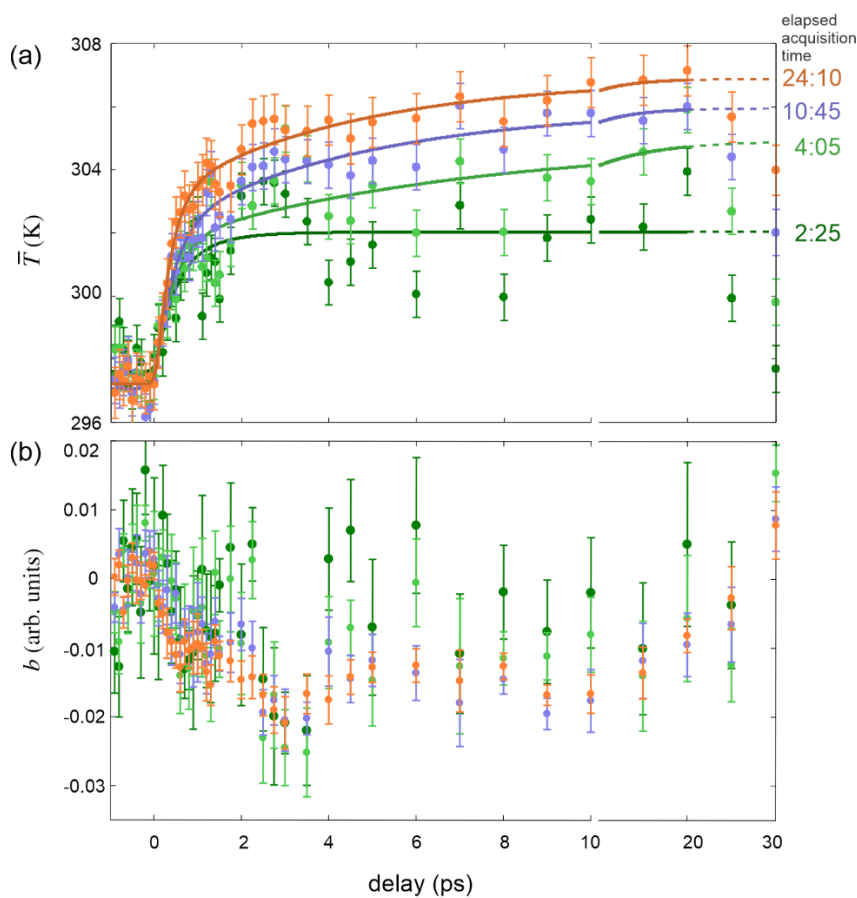

Figure S3 – (a) Mean lattice temperature and (b) the $b(t)$ parameter, as functions of pump-probe delay. The data were extracted in the same way as the data in Fig. 4a and 4b of the main text. The different curves are generated from data taken after different acquisition times up to 24h (the same times are used in both panels). The data in (a) emphasize that the sub-picosecond process is always present, but the slower process evolves during the measurement. The data in (b) demonstrate that the Q-dependent behavior is present and does not qualitatively change.

## 6. Friedel pair analysis

In this section we present background regarding the asymmetry analysis and deformation of the unit cell. In a rhombohedral system the size of a scattering vector reads:

$$Q^2(hkl) = \frac{(h^2 + k^2 + l^2)\sin^2\alpha + 2(hk + kl + hl)(\cos^2\alpha - \cos\alpha)}{a^2(1 - 3\cos^2\alpha + 2\cos^3\alpha)} \tag{S6}$$

From inspection we find that $\alpha = 90°$ recovers the cubic functional dependence. Asymmetry is defined in the main text as

$$\Lambda(hkl) = Q^2(hkl) - Q^2(\bar{h}\bar{k}\bar{l}) \tag{S7}$$

Inspecting Eq. (S6), we conclude that $\Lambda$ is always zero. However, an experiment probes a cut of reciprocal space given by Ewald's sphere, which is not flat. Therefore, reflections are not cut precisely through their center, but rather *near* their center, providing an *observed* maximum. This scenario is depicted in Figure S4 for a Friedel pair of the form $(hk0)$. In this depiction it is clear that both reflections in the pair acquire a small out of plane component $\delta$ *of the same sign*. Therefore, an assumed $\Lambda(hk0)$ is actually $\Lambda(h\ k\ |\delta|)$, which can be expressed as

$$\Lambda(hk0) = \frac{4\delta(k+h)\cos\alpha}{a^2(\cos\alpha - 1)(1 + 2\cos\alpha)} \equiv \delta a^{-2}(h+k)f(\alpha) \tag{S8}$$

When plugging in $\alpha = 0°$, we find $\Lambda(hk0) = 0$, so even if $\delta$ exists (experimentally it always does), <u>a non-zero $\Lambda$ can only occur if the system is not orthonormal</u>. This description is valid if $a = b$, as in the cubic case. In our experiment we do not identify any deviation from this.

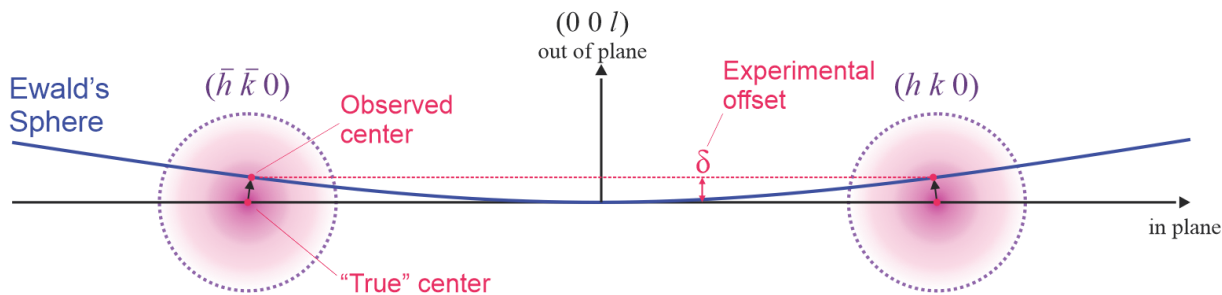

Figure S4 – sketch of the experimental observation of Friedel pair with $l = 0$. The curvature of Ewald's sphere causes both reflections to acquire a small out-of-plane component $l = \delta$.

In the following we present examples of $\Lambda$. Importantly, we note that asymmetry will occur in our data also due to experimental artifacts related to the experimental geometry (e.g. sample tilts). These effects would constitute a constant baseline for $\Lambda$. To avoid this, we consider only the *photoinduced changes* to $\Lambda$, and not the absolute quantity, by subtracting the baseline. The top panels of Figure S5 present $\Lambda(t)$ for six Friedel pairs along the two primary axes: $\pm(h\ 0\ 0)$ and $\pm(0\ k\ 0)$. Since the data here are based on the positions of Bragg peaks on our 2D detector, we avoid any possible artifacts associated with unit conversion by presenting in units of detector pixels instead of Å$^{-1}$. We find that dynamics observed for all six pairs along a given axis are similar, but are larger in magnitude as $h$ or $k$ grows.

To understand this, we reexamine Eq. (S8). Since $a = b = c$, photoinduced dynamics of $\Lambda$ occur only due to $\alpha$, the rhombohedral angle. Furthermore, the photoinduced change of $\Lambda$ should scale in magnitude with $(h + k)$. To test this, the lower panels of Figure S5 present the same data, divided by $h$ or $k$. We find that this causes the $\Lambda(t)$ curves to fall onto each other. This demonstrates that the photoinduced changes of $\Lambda$ that we observe are in good agreement with Eq. (S8), pointing to a change in the unit cell's shape.

We note that while the changes originate from the rhombohedral angle $\alpha$, the change in $\alpha$ may not be continuous. Within the observed time frame it could occur that $\alpha \neq \beta \neq \gamma$, in which case $f(\alpha)$ in Eq. (S8) would be $f(\alpha, \beta, \gamma)$.

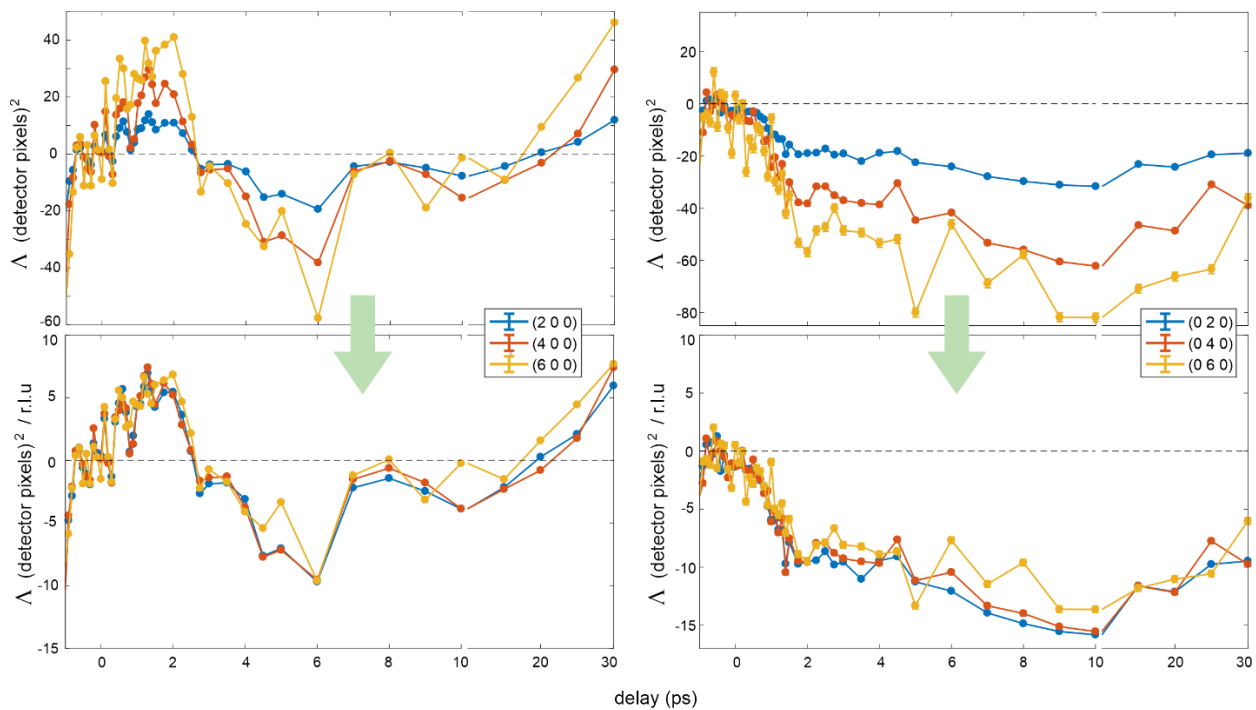

Figure S5 – Asymmetries $\Lambda$ in "unprocessed" units of detector pixels (squared), from reflections along the principle axes [010] (left) and [100] (right). The bottom row presents the same data, normalized by $(h + k)$.